\documentstyle[twoside,amsfonts,amsmath]{article}

\newtheorem{remark}{Remark}

\catcode`\@=11
\long\def\@makefntext#1{
\protect\noindent \hbox to 3.2pt {\hskip-.9pt  
$^{{\eightrm\@thefnmark}}$\hfil}#1\hfill}		

\def\@makefnmark{\hbox to 0pt{$^{\@thefnmark}$\hss}}	
	
\def\ps@myheadings{\let\@mkboth\@gobbletwo
\def\@oddhead{\hbox{}
\rightmark\hfil\eightrm\thepage}   
\def\@oddfoot{}\def\@evenhead{\eightrm\thepage\hfil
\leftmark\hbox{}}\def\@evenfoot{}
\def\sectionmark##1{}\def\subsectionmark##1{}}

\oddsidemargin=\evensidemargin
\newcounter{sectionc}\newcounter{subsectionc}\newcounter{subsubsectionc}
\renewcommand{\section}[1] {\vspace{12pt}\addtocounter{sectionc}{1} 
\setcounter{subsectionc}{0}\setcounter{subsubsectionc}{0}\noindent 
	{\tenbf\thesectionc. #1}\par\vspace{5pt}}
\renewcommand{\subsection}[1] {\vspace{12pt}\addtocounter{subsectionc}{1} 
\setcounter{subsubsectionc}{0}\noindent 
{\bf\thesectionc.\thesubsectionc. {\kern1pt \bfit #1}}\par\vspace{5pt}}
\renewcommand{\subsubsection}[1] {\vspace{12pt}\addtocounter{subsubsectionc}{1}
	\noindent{\tenrm\thesectionc.\thesubsectionc.\thesubsubsectionc.
	{\kern1pt \tenit #1}}\par\vspace{5pt}}
\newcommand{\nonumsection}[1] {\vspace{12pt}\noindent{\tenbf #1}
	\par\vspace{5pt}}

\newcounter{appendixc}
\newcounter{subappendixc}[appendixc]
\newcounter{subsubappendixc}[subappendixc]
\renewcommand{\thesubappendixc}{\Alph{appendixc}.\arabic{subappendixc}}
\renewcommand{\thesubsubappendixc}
	{\Alph{appendixc}.\arabic{subappendixc}.\arabic{subsubappendixc}}

\renewcommand{\appendix}[1] {\vspace{12pt}
        \refstepcounter{appendixc}
        \setcounter{figure}{0}
        \setcounter{table}{0}
        \setcounter{lemma}{0}
        \setcounter{theorem}{0}
        \setcounter{corollary}{0}
        \setcounter{definition}{0}
        \setcounter{equation}{0}
        \renewcommand{\thefigure}{\Alph{appendixc}.\arabic{figure}}
        \renewcommand{\thetable}{\Alph{appendixc}.\arabic{table}}
        \renewcommand{\theappendixc}{\Alph{appendixc}}
        \renewcommand{\thelemma}{\Alph{appendixc}.\arabic{lemma}}
        \renewcommand{\thetheorem}{\Alph{appendixc}.\arabic{theorem}}
        \renewcommand{\thedefinition}{\Alph{appendixc}.\arabic{definition}}
        \renewcommand{\thecorollary}{\Alph{appendixc}.\arabic{corollary}}
        \renewcommand{\theequation}{\Alph{appendixc}.\arabic{equation}}
        \noindent{\tenbf Appendix \theappendixc #1}\par\vspace{5pt}}
\newcommand{\subappendix}[1] {\vspace{12pt}
        \refstepcounter{subappendixc}
        \noindent{\bf Appendix \thesubappendixc. {\kern1pt \bfit #1}}
	\par\vspace{5pt}}
\newcommand{\subsubappendix}[1] {\vspace{12pt}
        \refstepcounter{subsubappendixc}
        \noindent{\rm Appendix \thesubsubappendixc. {\kern1pt \tenit #1}}
	\par\vspace{5pt}}

\newcommand{\textlineskip}{\baselineskip=13pt}
\newcommand{\smalllineskip}{\baselineskip=10pt}

\def\abstracts#1#2#3{{
	\centering{\begin{minipage}{4.5in}\footnotesize\baselineskip=10pt
	\parindent=0pt #1\par 
	\parindent=15pt #2\par
	\parindent=15pt #3
	\end{minipage}}\par}} 

\def\keywords#1{{
	\centering{\begin{minipage}{4.5in}\footnotesize\baselineskip=10pt
	{\footnotesize\it Keywords}\/: #1
	 \end{minipage}}\par}}

\renewenvironment{thebibliography}[1]
        {\frenchspacing
	 \ninerm\baselineskip=11pt
         \begin{list}{\arabic{enumi}.}
        {\usecounter{enumi}\setlength{\parsep}{0pt}     
	 \setlength{\leftmargin 12.7pt}{\rightmargin 0pt}
         \setlength{\itemsep}{0pt} \settowidth
	{\labelwidth}{#1.}\sloppy}}{\end{list}}

\newcounter{itemlistc}
\newcounter{romanlistc}
\newcounter{alphlistc}
\newcounter{arabiclistc}

\newcommand{\fcaption}[1]{
        \refstepcounter{figure}
        \setbox\@tempboxa = \hbox{\footnotesize Fig.~\thefigure. #1}
        \ifdim \wd\@tempboxa > 5in
           {\begin{center}
        \parbox{5in}{\footnotesize\smalllineskip Fig.~\thefigure. #1}
            \end{center}}
        \else
             {\begin{center}
             {\footnotesize Fig.~\thefigure. #1}
              \end{center}}
        \fi}

\newcommand{\tcaption}[1]{
        \refstepcounter{table}
        \setbox\@tempboxa = \hbox{\footnotesize Table~\thetable. #1}
        \ifdim \wd\@tempboxa > 5in
           {\begin{center}
        \parbox{5in}{\footnotesize\smalllineskip Table~\thetable. #1}
            \end{center}}
        \else
             {\begin{center}
             {\footnotesize Table~\thetable. #1}
              \end{center}}
        \fi}

\def\@citex[#1]#2{\if@filesw\immediate\write\@auxout
	{\string\citation{#2}}\fi
\def\@citea{}\@cite{\@for\@citeb:=#2\do
	{\@citea\def\@citea{,}\@ifundefined
	{b@\@citeb}{{\bf ?}\@warning
	{Citation `\@citeb' on page \thepage \space undefined}}
	{\csname b@\@citeb\endcsname}}}{#1}}

\newif\if@cghi
\def\cite{\@cghitrue\@ifnextchar [{\@tempswatrue
	\@citex}{\@tempswafalse\@citex[]}}
\def\citelow{\@cghifalse\@ifnextchar [{\@tempswatrue
	\@citex}{\@tempswafalse\@citex[]}}
\def\@cite#1#2{{$\null^{#1}$\if@tempswa\typeout
	{IJCGA warning: optional citation argument 
	ignored: `#2'} \fi}}

\def\pmb#1{\setbox0=\hbox{#1}
	\kern-.025em\copy0\kern-\wd0
	\kern.05em\copy0\kern-\wd0
	\kern-.025em\raise.0433em\box0}

\def\fnt#1#2{\footnotetext{\kern-.3em
	{$^{\mbox{\scriptsize #1}}$}{#2}}}

\def\fpage#1{\begingroup
\voffset=.3in
\thispagestyle{empty}\begin{table}[b]\centerline{\footnotesize #1}
	\end{table}\endgroup}

\def\runninghead#1#2{\pagestyle{myheadings}
\markboth{{\protect\footnotesize\it{\quad #1}}\hfill}
{\hfill{\protect\footnotesize\it{#2\quad}}}}
\font\tenrm=cmr10
\font\tenit=cmti10 
\font\tenbf=cmbx10
\font\bfit=cmbxti10 at 10pt
\font\ninerm=cmr9

\font\eightrm=cmr8

\newcommand{\proof}[1]{{\bf Proof.} #1 $\Box$.}

\def\FigName{figure}%
\newbox\captionbox
\long\def\@makecaption#1#2{%
  \ifx\FigName\@captype
    \vskip\abovecaptionskip
    \setbox\tempbox\hbox{{\figurecaptionfont #1\hskip1em #2}}
	\ifdim\wd\tempbox< 28pc
	\centerline{\box\tempbox}
	\else
	{\figurecaptionfont #1\hskip1em #2\par}
\fi\else
  	\setbox\tempbox\hbox{{\tablecaptionfont #1\hskip1em #2}}
 	\ifdim\wd\tempbox< 28pc 
	\centerline{\box\tempbox}
	\else
	{\tablecaptionfont #1\hskip1em #2\par}%
	\fi   
 \vskip\belowcaptionskip
 \fi}
\InputIfFileExists{psfig.sty}{\typeout{^^Jpsfig.sty 
inputed...ok}}{\typeout{^^JWarning: psfig.sty could be be found.^^J}}
\InputIfFileExists{epsfsafe.tex}{\typeout{^^Jepsfsafe.tex inputed...ok}}
			{\typeout{^^JWarning: epsfsafe.tex could not be 
found.^^J}}
\InputIfFileExists{epsfig.sty}{\typeout{^^Jepsfig.sty 
inputed...ok}}{\typeout{^^JWarning: epsfig.sty could not be found.^^J}}
\InputIfFileExists{epsf.sty}{\typeout{^^Jepsf.sty 
inputed...ok}}{\typeout{^^JWarning: epsf.sty could not be found.^^J}}%
\def\fps@figure{tbp}
\def\ftype@figure{1}
\def\ext@figure{lof}
\def\fnum@figure{Fig.\ \thefigure}
\def\qed{\hbox{${\vcenter{\vbox{	          
   \hrule height 0.4pt\hbox{\vrule width 0.4pt height 6pt
   \kern5pt\vrule width 0.4pt}\hrule height 0.4pt}}}$}}

\begin{document}

\runninghead{On Using Quantum Protocols to Detect Traffic Analysis} 
            {R. Steinwandt, D. Janzing, and Th. Beth}

\normalsize\textlineskip
\thispagestyle{empty}
\setcounter{page}{1}


\fpage{1}
\centerline{\bf
ON USING QUANTUM PROTOCOLS TO DETECT TRAFFIC ANALYSIS}
\vspace*{0.37truein}
\centerline{\footnotesize 
RAINER STEINWANDT, DOMINIK JANZING, and THOMAS BETH}
\vspace*{0.015truein}
\begin{center}{\footnotesize\it Institut f\"ur Algorithmen und Kognitive
Systeme, Fakult\"at f\"ur Informatik,

Am Fasanengarten 5, Universit\"at Karlsruhe (TH),

76131 Karlsruhe, Germany}
\end{center}
\baselineskip=10pt
\vspace*{0.225truein}

\vspace*{0.21truein}
\abstracts{
   We consider the problem of detecting whether an
   attacker measures the amount of traffic sent over a communication
   channel---possibly without extracting information about the
   transmitted data. A basic approach for designing a quantum protocol
   for detecting a perpetual traffic analysis of this kind is described.}{}{}

\vspace*{10pt}


\vspace*{1pt}\textlineskip	
\section{Introduction}

\noindent Within classical cryptography it is a well-known phenomenon that a 
communication
channel can be eavesdropped without affecting the transmitted data. In recent 
years
techniques have been developed which
exploit the quantum properties of the microphysical world in order to
deal with this problem. One of the most prominent among these methods is the
protocol for quantum key distribution described by Bennett and 
Brassard\cite{BB84}.
By means of such quantum protocols it is possible to ensure that an eavesdropper 
who tries
to get knowledge of the transmitted data can be detected with high probability.
However, an interesting aspect of eavesdropping which does not seem to be 
covered by
the quantum protocols suggested so far is traffic analysis: think of
an attacker who is only interested in analyzing the amount of traffic
sent over a channel, i.\,e., the attacker is only interested in
knowing how much data is transmitted and not
necessarily in reading the transmitted information itself.

If the communication channel is part of a network then one approach to thwart
such a traffic analysis is to conceal the identity of the recipient of the
transmitted data (cf., e.\,g., Rackoff and Simon\cite{RaSi93} and the references 
given
there).  In case of an ``isolated'' communication channel the situation is much 
worse---if the
existence of the channel cannot be kept secret (say by means of steganographic
techniques)
the classical approach to circumvent an analysis of the amount of meaningful 
data sent
over the channel is to keep the communication channel busy all the
time. This means that data is sent continuously, and the legitimate users of the
channel have to separate the relevant from the ``dummy'' data via a suitable 
(secret)
synchronization. No other solution to this problem seems to be known---e.\,g., 
Rackoff and Simon\cite[Section~1]{RaSi93} state ``In
other words, any secret regarding the total volume of information sent or 
received by a
party is purchased at the cost of the extra `dummy traffic' generated to 
disguise it.''

In this contribution we describe a simple quantum protocol which under
suitable assumptions enables the legitimate users of a quantum channel
to detect with high probability whether the traffic on a quantum
channel is analyzed---even if the attacker does not try to read the
transmitted data. We are not aware of a classical analogue of such a
procedure. However, we emphasize that in the present form, the protocol is not
suitable for practical cryptographic use,
as e.\,g., topics like authentication, noisy communication channels, or
efficiency are not taken into account. Nevertheless, we think the
described protocol to be of interest, as it might point out a new
application of quantum physical phenomena for cryptographic purposes.

Roughly speaking, the idea of the protocol is as follows:
as carrier of the information to be transmitted we think of using an inner 
degree of
freedom of some particle (like the polarization of a photon) or different types 
of particles for
representing zero and one.
Then to transmit a bit, first the wave function of the particle carrying the
information is split up into two parts, and one of these parts is
transmitted over the communication channel; the other part is kept
secret by the sender. At no time more than one wave package is present
on the communication channel.
The receiving party chooses at random whether the transmitted bit
is received or mirrored back to the sender---in the latter case the
corresponding bit is sent again in the next step.
If no attack takes place and a wave package is returned, then there will be
interference between a returned wave package and the
wave package retained by the sender. As an observation of
the channel causes a partial collapse of the wave function, a traffic
analysis destroys this interference.

\section{A Quantum Protocol for Detecting Traffic
Analysis}

\noindent Following the established terminology, the communicating parties will 
be
refered to as Alice and Bob  in the
sequel. Moreover, the attacker who is interested in
analyzing the traffic on the communication channel between Alice and
Bob will be called Tracy. We avoid the name Eve, as Tracy is not
necessarily interested in eavesdropping the channel, i.\,e., obtaining
the transmitted information as such. Instead, she may restrict her
interest on learning the number of transmitted bits.

\subsection{Statement of the protocol}

\noindent Let us suppose that Alice wants to send a message $m=m_1\dots m_{n}$ 
to
Bob (we assume the $m_i$'s to be individual bits,
but one could as well use a coarser subdivision of $m$). Moreover, Alice
would like to know whether
Tracy is analyzing the communication channel while
$m$ is being sent to Bob.
If $n$ is not too small, then informally the technical apparatus and
a protocol for doing so
can be described as follows (for a more formal treatment see
Section~2.2):

\vbox{\noindent{\bf Technical apparatus}
\begin{itemize}
   \item Alice has a source for producing single particles,
   and she can encode a bit value in an inner degree
   of freedom of such a particle (as an example we may think of the
   polarization of a photon). {Alternatively one can also
   think of using different types of particles for encoding zero
   and one}.
   Moreover, Alice has a 50:50 beam
   splitter ${\cal B}_1$ which splits a particle into two wave packages without
   affecting this inner degree of freedom, and she is able to
   store {one} such wave package in her laboratory.
   Finally, she has a 50:50
   beam splitter ${\cal B}_2$ where a wave package received on the channel and
   a wave package stored in her laboratory can meet in such a way that
   in one branch constructive and in the other branch destructive
   interference occurs. By means of a detector $\cal C$ in the ``constructive
   branch'' and a detector $\cal D$ in the ``destructive branch''
   Alice can decide
   whether a particle is present in one of the branches.

   \item Bob needs a switchable device which either mirrors
   a wave package on the channel back to Alice without
   detecting the presence of the package, or reads out the inner
   degree of freedom of a particle on the channel without returning anything.
\end{itemize}}

\noindent {\bf Initialization}

\noindent We assume that Alice and Bob have agreed upon a time $t_1$ where the
transmission of Alice is to begin. By $\Delta\in{\mathbb R}$ we denote
{the time required by a
particle for passing from Alice to Bob back to Alice}. Alice and Bob
can use $\Delta$ to synchronize their transmissions in such a way that
there is at no time more than one wave package present on the channel,
{and Alice can make use of $\Delta$ to carry out her
interference experiments correctly}. Finally,
Alice initializes a variable $a\leftarrow0$
for counting the number of transmitted wave packages, and Bob
initializes a boolean variable $b\leftarrow\mbox{true}$.\medskip

\noindent{\bf Transmission protocol}

\noindent For $i\leftarrow[1,\ldots,n]$ Alice proceeds as follows to transmit  
bit $m_i$
to Bob:

\begin{enumerate}
   \item Alice produces a single particle $P$ and
   encodes $m_i$ in an inner degree of freedom of $P$.
   Then she
   passes $P$ through the beam splitter ${\cal B}_1$ and
   retains one of the two resulting wave packages in her
   private laboratory. The other wave package is transmitted to Bob
   over the communcication channel at time $t_i=t_1+a\cdot
   \Delta$. Then Alice sets $a\leftarrow a+1$.

   \item If $b=\mbox{true}$ then Bob selects
           $r\in\{\mbox{true},\mbox{false}\}$ at random.
           Otherwise, i.\,e., for
           $b=\mbox{false}$, the value of $r$ remains unchanged.

   \item If $r=\mbox{true}$ then Bob mirrors the wave package back to
   Alice. Otherwise he
   reads out the inner degree of freedom of the potentially present
   particle on the communication channel.

   \item Bob sets $(b,r)\leftarrow(\mbox{not } b,\mbox{not }r)$.

   \item  Using the beam splitter ${\cal B}_2$ Alice brings the wave
          package potentially returned by Bob and the wave package
          stored in her laboratory into interference.
          Let $C_j$ resp. $D_j$ ($1\le j\le a$) be the random variable
          describing the number of particles ($0$ or $1$) detected in
          detector ${\cal C}$ resp. $\cal D$ in Alice's
          $j^{\rm th}$ interference experiment.

   \item If $a$ is even then Alice tests the following hypothesis:

   the vector valued random
   variables $(C_{2j-1}, C_{2j}, D_{2j-1}, D_{2j})$ (where $1\le j\le a/2$)
   are identically independently distributed with the probability
   distribution
   $$\renewcommand{\arraystretch}{1.2}\begin{array}{rlcclccl}
      p\bigl(\hspace*{-5pt}&(C_{2j-1}, C_{2j}, D_{2j-1}, 
D_{2j})&\hspace*{-6pt}=\hspace*{-6pt}&(1,1,0,0)&\hspace*{-5pt}\bigr)&=&1/4,&\\
      p\bigl(\hspace*{-5pt}&(C_{2j-1}, C_{2j}, D_{2j-1}, 
D_{2j})&\hspace*{-6pt}=\hspace*{-6pt}&(1,0,0,0)&\hspace*{-5pt}\bigr)&=&{1}/{4},&
\\
      p\bigl(\hspace*{-5pt}&(C_{2j-1}, C_{2j}, D_{2j-1}, 
D_{2j})&\hspace*{-6pt}=\hspace*{-6pt}&(0,1,0,0)&\hspace*{-5pt}\bigr)&=&{1}/{4},&
\\
      p\bigl(\hspace*{-5pt}&(C_{2j-1}, C_{2j}, D_{2j-1}, 
D_{2j})&\hspace*{-6pt}=\hspace*{-6pt}&(1,0,0,1)&\hspace*{-5pt}\bigr)&=&{1}/{8},&
\\
      p\bigl(\hspace*{-5pt}&(C_{2j-1}, C_{2j}, D_{2j-1}, 
D_{2j})&\hspace*{-6pt}=\hspace*{-6pt}&(0,1,1,0)&\hspace*{-5pt}\bigr)&=&{1}/{8},&
\\
      p\bigl(\hspace*{-5pt}&(C_{2j-1}, C_{2j}, D_{2j-1}, 
D_{2j})&\hspace*{-6pt}=\hspace*{-6pt}&x&\hspace*{-5pt}\bigr)&=&0&\mbox{for all
      other $x$}.
   \end{array}$$

   If this hypothesis has to be rejected with high
   probability then Tracy is assumed to analyze the communication
   channel and the protocol is aborted.

   \item  If $C_{a}+D_a=0$ then Alice assumes that the transmission
   of bit $m_i$ is complete, i.\,e., that Bob has succeeded
   in receiving $m_i$. Otherwise, i.\,e., if $C_{a}+D_{a}\ne 0$,
   Alice goes back to Step~1---thereby retransmitting $m_i$.
\end{enumerate}
Before explaining and analyzing the above protocol in
more detail we would like to emphasize again that in the above simple form
the protocol is not suitable for practical cryptographic use. In
particular it does not take the problem of authentication into account;
also it is assumed that the communication channel is perfect. Similarly, for
ease of presentation, in Step~6 of the above protocol, Alice does not
check whether a detected particle indeed encodes the bit $m_i$.

\subsection{Explanation and analysis of the protocol}

\noindent To explain the protocol we introduce some notation. For sake of
simplicity for the moment we restrict ourselves to an informal explanation of
the individual steps of the protocol; a more rigorous
treatment is postponed until later.

By ${\cal G}$ we denote the Hilbert space corresponding to the particle's inner 
degrees of
freedom and by ${\cal H}$ the Hilbert space spanned by the following
three basis states:
\begin{enumerate}
   \item $|a\rangle$: the particle $P$ is in Alice's laboratory
   \item $|c\rangle$: the particle $P$ is in the communication channel
   \item $|b\rangle$: the particle $P$ is in Bob's laboratory
\end{enumerate}

In the first step of the protocol Alice prepares the state
\begin{equation}\label{equ:thestate}
  \frac{1}{\sqrt{2}}(|a\rangle +|c\rangle)\otimes|m_i\rangle
\end{equation}
 where $|m_i\rangle\in{\cal G}$ stands for the inner state representing
the bit $m_i$.

In the Steps~2--4 the role of Bob's boolean variables
$b$ and $r$ is as follows: the variable $b$ is used to group the
transmitted wave packages of Alice into pairs in such a way that for
$j\in{\mathbb N}$ arbitrary either the $(2j-1)^{\rm st}$ or the $(2j)^{\rm th}$
wave package (but never both of them) is mirrored back to Alice:
$b=\mbox{true}$ resp. $b=\mbox{false}$ means that currently a ``first''
($(2j-1)^{\rm st}$) resp.
``second'' ($(2j)^{\rm th}$) element of a pair is processed by Bob. The
random choice of $r$ in Step~2 is used to decide at random whether the
first or second wave package of the current pair is mirrored back.
Note that the decision whether to return the first or second package is made
independently for each individual pair, and
we assume that mirroring back the wave package does not alter
the state~(\ref{equ:thestate}).

In Step~3 if $r={\rm false}$ then Bob attempts to read out the inner
degree of freedom of the potentially present particle. For this the
wave package enters his laboratory which translates into transforming the
state~(\ref{equ:thestate}) into
\begin{equation}\label{equ:thebobstate}
  \frac{1}{\sqrt{2}}(|a\rangle +|b\rangle)\otimes|m_i\rangle.
\end{equation}
Reading out the inner degree of freedom then means to perform
a measurement on the state~(\ref{equ:thebobstate}).
This measurement is described by the following three projectors:
\begin{enumerate}
  \item $Q_0$: Projection on the one-dimensional subspace spanned by
               $|b\rangle\otimes|0\rangle\in{\cal H}\otimes{\cal G}$---Bob 
receives $m_i=0$.
  \item $Q_1$: Projection on the one-dimensional subspace spanned by
               $|b\rangle\otimes|1\rangle\in{\cal H}\otimes{\cal G}$---Bob 
receives $m_i=1$.
  \item $Q_\epsilon:=({\boldsymbol{1}}-|b\rangle\langle 
b|)\otimes{\boldsymbol{1}}$---Bob does not detect a
               particle at all.
\end{enumerate}
In the sequel the corresponding measurement outcomes are denoted by
$0$, $1$, and $\epsilon$.

Finally, to see why Alice's hypothesis in Step~6 is not rejected with high
probability reduces to computing the five non-zero probabilities occuring in
the hypothesis:

\begin{itemize}
 \item $p\bigl((C_{2j-1}, C_{2j}, D_{2j-1},
 D_{2j})=(1,1,0,0)\bigr)=1/4$: w.\,l.\,o.\,g. we assume
 that Bob measures the bit transmitted in the $(2j-1)^{\rm st}$
 transmission. Then the result $C_{2j-1}=1$ is only
 possible if Bob's measurement yields the result $\epsilon$.
 In this case we have $C_{2j-1}=1$ with probability $1/2$. Since the
 result $\epsilon$ has
 probability $1/2$ as well, we obtain $p\bigl((C_{2j-1}, C_{2j}, D_{2j-1},
 D_{2j})=(1,1,0,0)\bigr)=1/4$ as required.  

 \item $p\bigl((C_{2j-1}, C_{2j}, D_{2j-1},
 D_{2j})=(1,0,0,0)\bigr)=1/4$: from $C_{2j}=0$ we conclude that Bob
 measured the ${2j}^{\rm th}$ transmission. This choice occurs with
 probability $1/2$ and also guarantees $C_{2j-1}=1$ and $D_{2j-1}=0$. 
 As $C_{2j}=D_{2j}=0$, the result of Bob's
 measurement is different from $\epsilon$; the latter event has probability
 $1/2$. So in summary we get $p\bigl((C_{2j-1}, C_{2j}, D_{2j-1},
 D_{2j})=(1,0,0,0)\bigr)=1/4$.

 \item $p\bigl((C_{2j-1}, C_{2j}, D_{2j-1},
 D_{2j})=(0,1,0,0)\bigr)=1/4$: the same argument as in the previous
 case with $2j$ and $2j-1$ interchanged.

 \item $p\bigl((C_{2j-1}, C_{2j}, D_{2j-1},
D_{2j})=(1,0,0,1)\bigr)=1/8$: from $C_{2j}=0$ we conclude that Bob
 measured the ${2j}^{\rm th}$ transmission. This choice occurs with
 probability $1/2$ and also guarentees $C_{2j-1}=1$ and $D_{2j-1}=0$.
 As $D_{2j}=1$ the result of Bob's measurement is $\epsilon$; the latter
 event has probability $1/2$. If Bob does not detect a particle
 (i.\,e., the result of his measurement is $\epsilon$) the
probability
 that $\cal C$ (resp. $\cal D$) detects a particle, is $1/2$.
 Consequently, we obtain $p\bigl((C_{2j-1}, C_{2j}, D_{2j-1},
D_{2j})=(1,0,0,1)\bigr)=1/8$.

\item $p\bigl((C_{2j-1}, C_{2j}, D_{2j-1},
D_{2j})=(0,1,1,0)\bigr)=1/8$: the same argument as in the previous
 case with $2j$ and $2j-1$ interchanged.
\end{itemize}
\noindent As the probabilities of these cases sum up to $1$ already, no other
values of the tuple $(C_{2j-1}, C_{2j}, D_{2j-1}, D_{2j})$ can occur.

Next, we show that measurements of Tracy which are appropriate
for detecting whether bits have been transmitted or not change the statistics
in Step~6 of the above transmission protocol. For this
we use a field-theoretic formulation of
the quantum physical situation.
The particles used in the protocol can be Bosons or Fermions,
here we restrict our attention to Bosons. Hence  the
field-theoretic description will represent the system in a symmetric
Fock space\cite{Ja68}.
If an arbitrary particle is described in a Hilbert space $\cal L$ the
corresponding Bose field is described in the symmetric
Fock space ${\cal F}^+({\cal L})$.
Here we take ${\cal L}:=L^2({\mathbb R}^3)$, the set of square
integrable functions (cf. the work of Prugove{\v c}ki\cite{Pr71}, for instance) 
as one-particle space.

Let $A\subseteq{\mathbb R}^3$ be the area of Alice's laboratory and
$X:={\mathbb R}^3$ its set-theoretic complement. Moreover, for
$R\subseteq {\mathbb R}^3$ we set $${\cal L}_R:=\{f\in{\cal L}|\ \forall
y\in({\mathbb R}^3\setminus R): f(y)=0\}.$$
So in particular we have ${\cal L}={\cal L}_A\oplus {\cal L}_X$. Moreover,
the algebra of operators acting on ${\cal F}^+({\cal L})$
can be split up into a tensor product of those which can be measured
in Alice's laboratory and those which can be measured outside
due to the formula
\begin{equation}\label{formeltenszer}
{\cal F}^+({\cal L}_A\oplus {\cal L}_X)={\cal F}^+({\cal
L}_A)\otimes{\cal F}^+({\cal L}_X).
\end{equation}
To express this more precisely, we can make use of the concept of a
Positive Operator Valued Measure (POVM)\cite{Da76}. A POVM
is a family $(a_i)_{i\in I}$ of positive operators with $\sum_{i\in
I} 
 a_i={\boldsymbol{1}}$; it describes the most general quantum mechanical
measurement.
As Tracy does not have access to Alice's laboratory,
every measurement which she can perform is a POVM
of the form
$(a_i)=({\boldsymbol{1}}\otimes b_i)$ where $b_i$ is an arbitrary positive
operator of the right-hand component of the tensor
product~(\ref{formeltenszer}).$^{\rm a}$\footnotetext{$^{\rm a}$This is a 
locality
assumption for the laws of physics: the operators which represent
measurements performed inside a certain area act trivially on the Hilbert space
corresponding to the fields outside the area. This is formalized by 
Haag\cite{Ha92} in a rather axiomatic approach.}

Now we distinguish between the two possible cases:
\begin{enumerate}
\item Alice transmits a particle to Bob: in this case the state vector
of the field is contained in the one-particle
subspace of ${\cal L}$. In this subspace the state can be 
described by a wave function $|\psi\rangle \in {\cal L}$. 

At some time $T_1$ one part of the wave function leaves Alice's Laboratory
and returns at a time $T_2$. Hence
 for $t\in [T_1,T_2]$ the state $|\psi(t)\rangle$ splits into a sum
\[
|\psi (t)\rangle:=\frac{1}{\sqrt{2}}(|\psi_1(t)\rangle +|\psi_2(t)\rangle)
\]
where the $|\psi_i(t)\rangle$ 
describe the outputs of the two branches of Alice's beam splitter
${\cal B}_1$: $\psi_1$ is the part of the wave function which is to
remain in Alice's laboratory, and $\psi_2$ is the part of the wave
function which is to be transmitted over the communication channel.
If we consider $|\psi(t)\rangle$ canonically as an element of the 
Fock space ${\cal F}^+({\cal L}_A)\otimes {\cal F}^+({\cal L}_X)$
we obtain
\[
|\psi(t)\rangle =|\psi_A(t)\rangle \otimes |0\rangle +|0\rangle \otimes 
|\psi_X(t)\rangle,
\]
where $|0\rangle$ denotes the $0$-particle state in ${\cal F}^+({\cal L}_A)$ and
${\cal F}^+({\cal L}_X)$, respectively.  
Since at no time more than ``half of the wave function'' is  outside the 
laboratory, 
 the norm of $|\psi_X(t)\rangle$ is not greater than $1/\sqrt{2}$ at 
any time $t$.

\item Alice does not transmit a particle to Bob:
in this case the state of the quantum field is for every time $t$ 
given by the vector
$|\phi (t)\rangle=|0\rangle\otimes |0\rangle$, i.e., the $0$-particle state
of ${\cal F}^+({\cal L}_A)\otimes {\cal F}^+({\cal L}_X)$.
\end{enumerate}

Tracy's measurement can only distinguish between the states $|\phi (t)\rangle$
and $|\psi (t)\rangle$ if the corresponding POVM $(a_i)$ contains
an operator $a_i$ with the property 
\begin{equation}\label{neq}
\langle \psi (t)| a_i |\psi (t)\rangle \neq \langle \phi (t)|a_i|\phi
 (t)\rangle.
\end{equation}
On the other hand,
a measurement apparatus can only be non-disturbing if   
the measured states are eigenvectors of every $a_i$ of the
 POVM$^{\rm b}$\footnotetext{$^{\rm b}$In Kraus' work\cite{Kr93}  one can find 
equations describing the connection between
the POVM and the corresponding effect of the measurement on the state.} 
.
Hence Tracy's attack has to be a measurement with the property that
 $|\psi(t)\rangle$
 is an eigenvector of each $a_i={\boldsymbol{1}}\otimes b_i$. One checks
easily that this can only be the case if
$|0\rangle$ and $|\psi_X(t)\rangle$ are eigenvectors of $b_i$ with the same 
eigenvalue:
\begin{remark}\label{rem:rem1}
With the above notation $|\psi(t)\rangle$
 can only be an eigenvector of each $a_i={\boldsymbol{1}}\otimes b_i$ if
 $|0\rangle$ and $|\psi_X(t)\rangle$ are eigenvectors of $b_i$ with
 the same eigenvalue.
\end{remark}
\proof{
The two components $|\psi_A(t)\rangle\otimes|0\rangle$ and
$|0\rangle\otimes|\psi_X(t)\rangle$ of $|\psi(t)\rangle$ are contained in the
mutually orthogonal vector spaces
$|\psi_A\rangle\otimes{\cal F}^+({\cal L}_X)$ and
$|0\rangle\otimes{\cal F}^+({\cal L}_X)$, respectively. Both of
these vector spaces are invariant under ${\boldsymbol{1}}\otimes b_i$, and
hence the claim follows.}

Remark~\ref{rem:rem1} implies that the observable ${\boldsymbol{1}}\otimes b_i$ 
cannot distinguish between the state $|0\rangle$ and the state
$|\psi(t)\rangle$ in the sense that the right-hand and the left-hand side of
equation~(\ref{neq}) coincide.

So assuming that whenever the part of the wave function outside Alice's 
laboratory
is modified then there can be no perfect destructive (constructive)
interference in the branch corresponding to Alice's detector $\cal D$
($\cal C$), Tracy's traffic analysis also modifies the statistics in
Step~6 of the transmission protocol. Qualitatively, this argument also
holds for joint attacks (cf., e.\,g., Biham and Mor\cite{BiMo97}): as at no time 
more
than one particle is present on the channel, such an attack had to
access single particles in order to store quantum information for a
later joint measurement on the memory.

A quantitative analysis of the information-disturbance
trade-off\cite{Fu96} had to  
make use of the fact that there is no time at which
the norm of the part outside the laboratory is greater than
$1/\sqrt{2}$. However, such an analysis is beyond the scope of this
paper, and we do not pursuit this topic any further here.

\section{Conclusions}

\noindent We have demonstrated (without giving a quantitative analysis) that in
principle quantum mechanical phenomena can be used to detect a
perpetual traffic analysis. In other words, it is
possible to detect an attacker who perpetually measures the amount of
traffic on a communication channel even in the case that the attacker
does not read the transmitted data itself.


\nonumsection{References}

\end{document}